\documentclass[a4paper,%
   a4paper,%
   BCOR12mm,%
   12pt,%
   abstracton,%
   pointednumbers,%
   tablecaptionabove,%
   footinclude,%
   halfparskip,%
   normalheadings,%
   ]{scrartcl}

\usepackage{epsfig}
\usepackage[square,comma,numbers,sort&compress]{natbib}
\usepackage{multirow}

\newcommand{\UPLB}{University of the Philippines Los Ba\~{n}os}
\newcommand{\PERCEPT}{\mathcal{P}}
\newcommand{\SOCLOAF}{\mathrm{SocialLoaf}}
\newcommand{\SELFSL}{\mathrm{SelfSL}}
\newcommand{\GROUPSL}{\mathrm{GroupSL}}

\newcommand{\TASKVIS}{\mathrm{TaskVis}}
\newcommand{\JUSTICE}{\mathrm{Justice}}
\newcommand{\CONTRIB}{\mathrm{Contrib}}
\newcommand{\DOMINANCE}{\mathrm{Dominance}}
\newcommand{\SUCKER}{\mathrm{Sucker}}

\newcommand{\FEMALE}{\mathrm{Female}}
\newcommand{\MALE}{\mathrm{Male}}

\newtheorem{hypothesis}{Hypothesis}

\begin{document}
\title{\normalsize Social Loafing Among Members of\\Undergraduate Software Engineering Groups:\\Persistence of Perception Seven Years After}
\author{\normalsize Reginald Neil C. Recario, Marie Betel B. de Robles,\\ \normalsize Kristine Elaine P. Bautista, and Jaderick P. Pabico\\
 \normalsize Institute of Computer Science, \UPLB}
\date{}
\maketitle

\begin{abstract}
Social Loafing is the term used by social scientists to describe the tendency of an individual to decrease efforts when working in a group compared to when working alone. This behavioral tendency of a member has been known to negatively affect the performance of groups. Researchers, however, found out that there are various precursors to social loafing. To be able to design better computer science courses, we would like to find out if social loafing and its precursors exist in undergraduate groups that were organized for solving computing problems. We found out from 239 students in 2008 that the precursors task visibility, distributive justice, and intrinsic task involvement were negatively associated with social loafing while dominance, aggression and sucker effect each were positively correlated.


Seven years after, we surveyed 169 undergraduate students who are enrolled in various courses. They were members of software engineering groups formed to solve various real-world computational problems by implementing software projects as part of the requirements of the course. This time, our analysis show that task visibility is negatively associated with social loafing while contributions, dominance, aggression and sucker effect are positively correlated. 

We further found out that perception of social loafing exists and still persists among members of computer programming groups. Compared to our 2008 analysis, we provide in this paper detailed analysis based on demographic parameters such as gender, course taken, age group, type of residence (urban or rural), and region of residence. The implication of this result is that aside from the usual problems that an instructor faces in teaching software engineering-related courses, the presence of social loafing also adds to the impediment of teaching effectiveness. Thus, it is imperative that instructors and course designers consider the implications associated with social loafing when designing group projects.

\end{abstract}

\section{Introduction}
The management of courses that teach construction of real-world solutions to problems via cooperation and collaboration, such as the development of software projects, allow students to work in groups. This is how most higher education institutions (HEIs) in the Philippines would prepare their students to the working conditions of the real-world, albeit in a simulated way. The group dynamics is designed so that students actively participate in the construction and sharing of knowledge and experience, enhance personal problem-solving skills through teamwork, and learn valuable lessons regarding group communications that can be used later in real-world work environment~\citep{Becker98, Black02, Haythornthwaite06}. In most software development and engineering courses, the group work technique is employed by instructors in courses that teach the management and control of software production groups. However, one of the difficulties in assessing the output of students in this kind of training technique is that the individual contribution is sometimes indistinguishable from the group's output~\citep{Williams81}. One possible explanation for this is the difficulty of tracking the progress of individual student, compounded by the absence of proper tools and technology available and accessible to the instructor. During group work, when a student perceives that her\footnote{Please note that we used the female gender as our writing style only and would mean either or both genders.} individual inputs to the group will not be given due recognition, either by her group mates or by her instructor, her motivation to contribute to the group will be diminished~\citep{Jones84}. The student's motivation to perform her best is not affected anymore by either the benefits of claiming high grades due to high levels of effort, or the penalties of getting low grades due to low levels of effort in contributing to the group~\citep{Jones84}. Because of this individual perception, the group's productivity diminishes. 

One of the reasons for productivity losses in groups is an individual's behavioral tendency called {\em social loafing}~\citep{Latane79}. Social loafing is the tendency of a group member to decrease her individual effort when working in a group compared to when she is working alone~\citep{Williams91}. The opposite of social loafing is termed {\em social facilitation}, which is the tendency of an individual to exert more effort in the presence of others than alone~\citep{Cook01}. Studies on social loafing can be traced as far back as 1913 in the work of Max Ringelmann, a French agricultural engineer. Ringelmann studied the efficiency of animals, men, and machines in various agricultural applications where he observed that there was a decrease in overall performance of the group when the number of members was increased. His observation was later termed in his honor as the {\em Ringelmann Effect}. In another Ringelmann's experiment involving men who were pulling a crank to provide manual power to a flour mill, he observed that as more men were added into the task, each man began to rely on his neighbors to perform the desired output. He further observed that some men became content to let their hand  follow the crank while some went as far as letting the crank pull their hands~\citep{Kravitz86}. The former behavioral phenomenon was later termed social loafing while the latter was termed free riding. Free riding is technically defined as the action of an individual who share the benefits of the group and yet did not spend any amount of effort~\citep{Albanese85}. 

Studies show that social loafing has been occurring in various group tasks due to the perceived behavioral factors of the group members. However, these studies have been exclusively performed under controlled environment for easy measurement of the variables. Other studies have been conducted outside the laboratory environments, but under real-world job settings and cultures different from that of the Philippines (see for example~\citet{Earley89} and~\citet{Harkins80}). Seven year ago, a study on undergraduate groups was performed to increase the researchers' understanding and awareness of social loafing as it occurs in the classroom settings in a Philippines HEI~\citep{Pabico08}. In that study, the researchers found out that the perception of social loafing and other behavioral factors exist in the minds of students who are members of undergraduate computer programming groups. These behaviors were known to be precursors to social loafing. They further found out that social loafing is negatively correlated with task visibility, distributive justice, and intrinsic task involvement, while it is  positively correlated with dominance, aggression and sucker effect. 

Seven year after, this research study extends the work of~\citet{Pabico08} by looking deeper into the influence of local Filipino demography on social loafing under the classroom settings. Specifically, we wanted to find out if gender, age, geographical location, and location type (rural or urban) have an effect on the perception of social loafing by students. This is a very important issue because personalities of group members are usually associated with gender, age, geographic location and the type of location where the member grew up. Aside from the individual personalities, instructors must also consider the profile of the group members in designing the group's composition, such that the objectives of the course is met through optimal (or near-optimal) group dynamics.

The rest of the article is organized as follows. The following subsections discuss the antecedent behaviors to social loafing, where each discussion leads to the development of our hypotheses for this research. In Section~\ref{sec:method}, we discuss briefly the demography of the participants and the survey that we conducted. We present the results in Section~\ref{sec:result}. We finally conclude with a short discussion on the implications of this research in Section~\ref{sec:conclude}.

\subsection{Personal Degree of Social Loafing}

Perceived social loafing is the term for the belief of an individual that her co-members are social loafing~\citep{Comer95}. In our research, as in the work of~\citet{Pabico08}, we only measured the perception of a group member, not the actual output of the member perceived to be social loafing. The reason for this is that there is a possibility that the one perceived to be social loafing during classwork may actually struggle with the assigned concept, spend many hours of personal effort, learn a lot in the process, and yet contribute less than the others to the output of the group. Whether or not social loafing is actually occurring, our research only measured the perception following the methodology set forth by the research of~\citet{Mulvey98}, the same methodology used by~\citet{Pabico08}. We assumed  that group members will base their action on the perceived behaviors of fellow members.

In common group work in a classroom setting, individuals may actually learn but each member may perceive unequal effort. Once a member perceive that some member are maybe either social facilitating  or social loafing, it may affect her personal motivation to contribute. The act of group members carrying a free rider or social loafer has been termed {\em playing a sucker role}. Engaging in social loafing to avoid playing the sucker role is termed {\em playing the sucker effect}~\citep{Kerr83}. The work of~\citet{Pabico08} shows that social loafing exists in classrooms under Philippine settings. We claim here that social loafing still persists among students in our subject Philippine HEI despite the efforts of instructors to avoid such behavior in the classroom and laboratory settings. Thus, we hypothesize that:

\begin{hypothesis}
Social loafing is perceived to exist in undergraduate software engineering groups in the Philippines.\label{hyp:socloaf}
\end{hypothesis}

Social loafing exists if a student will report that either she, herself, is engaged in social loafing ($\SELFSL$), or that she perceives that her groupmates are engaged in social loafing ($\GROUPSL$). Thus, social loafing ($\SOCLOAF$) exists if the perception $\PERCEPT$ of $\SELFSL$ or $\PERCEPT$ of $\GROUPSL$ exists. Mathematically, that is: $\SOCLOAF = \PERCEPT_\SELFSL + \PERCEPT_\GROUPSL$.

\begin{hypothesis}
Sucker effect ($\SUCKER$) is a positive precursor to the perception of social loafing in undergraduate computer programming groups.\label{hyp:sucker}
\end{hypothesis}

Mathematically, $\PERCEPT_\SUCKER \propto \PERCEPT_\SOCLOAF$.

\subsection{Individual Task Visibility}

The belief that the class instructor is observing and tracking each student's input to the group is termed as {\em perceived task visibility}~\citep{Kidwell93}. If the group's assigned tasks are interdependent, the individual's perceived task visibility will decrease because tracking the individual contribution will be very difficult~\citep{Jones84}, specifically in the absence of specialized tools and appropriate technologies. When an individual's inputs become indistinguishable from the group, the individual becomes unable to associate her personal input and claim the benefits associated with the effort~{Jones84}. Moreover, an industrious member may feel inequity and decide to social loaf if she works with other members who do not suffer the consequences of not sufficiently contributing to the group. On the other hand, she who do not fully contribute may also social loaf because she may perceive that her inputs are not critical to the group's success~\citep{Karau93}. Further, she may also perceive an inequitable relationship~\citep{Walster73}, believe that benefits of social loafing outweigh the cost of the penalty~\citep{Murphy03}, or is intentionally free riding. Thus, we have state our hypothesis:

\begin{hypothesis}
Task visibility ($\TASKVIS$) is a negative precursor to the perception of social loafing in undergraduate computer programming groups.\label{hyp:taskvis}
\end{hypothesis}

Mathematically, the above hypothesis is $\PERCEPT_\TASKVIS \propto -\PERCEPT_\SOCLOAF$.

\subsection{Just Grade Distribution}

{\em Perceived distributive justice} is an individual's perception of the justified distribution of grades among group members~\citep{Liden04}. {\em Procedural justice}~\citep{Greenberg90}, on the other hand, is the individual's perceived fairness of the procedures and policies used to compute for the grades. When participating in group tasks, the achievement of a student may be influenced by her perception of procedural and distributive justice set forth by the instructor. A student might reduce her effort if there is perception of unfair equity of grade distribution~\citep{Kidwell93}. Researchers report that procedural justice and social loafing are significantly correlated (see for example~\citet{Liden04} and~\citet{Karau93}), such that the student's perception of fairness in the procedure for grade distribution may influence the student's effort on group tasks. Thus, 

\begin{hypothesis}
The perceived distributive justice ($\JUSTICE$) is a negative precursor to social loafing in undergraduate computer programming groups.\label{hyp:justice}
\end{hypothesis}

Hypothesis~\ref{hyp:justice} is thus, $\PERCEPT_\JUSTICE \propto -\PERCEPT_\SOCLOAF$.

\subsection{Dominance and Aggression}

Instructors usually consider the personalities of the group members in designing the group's composition. Group members with stronger personality usually dominate the group, thereby negatively affecting the group dynamics~\citep{Palloff03}. Subdued members perceived themselves to be dominated, intimidated, or harassed and eventually resort to social loafing~\citep{Michaelsen97}. Hence,

\begin{hypothesis}
Dominance and aggression ($\DOMINANCE$) are positive precursors to social loafing in undergraduate computer programming groups.\label{hyp:dominance}
\end{hypothesis}

Abstractlly, $\PERCEPT_\DOMINANCE \propto \PERCEPT_\SOCLOAF$.

\subsection{Individual Contribution}

A group member will likely exert extraordinary effort if she perceives that her individual effort within the group is meaningful~\citep{Karau93}. In a divided task, a student who was assigned the easy subtask may feel that she is being prejudiced and believe that her full effort is not required for the group's success~\citep{Liden04}. Similarly, if the member's inputs are highly integrated into the group's output while the corresponding grades are distributed accordingly, the individual motivation may also be affected negatively~\citep{Lawler71}. Thus,

\begin{hypothesis}
The perceived individual contribution ($\CONTRIB$) is a negative precursor to social loafing in undergraduate computer programming groups.\label{hyp:contrib}
\end{hypothesis}

The following mathematical expression abstractly captures the above hypothesis: $\PERCEPT_\CONTRIB \propto -\PERCEPT_\SOCLOAF$.

\section{Methodology}\label{sec:method}

\subsection{Demography of Participants}

We surveyed 169 undergraduate students enrolled in three different courses at UPLB during the Second Semester of Academic Year (AY) 2014-2015 and First semester of AY 2015-2016. The participants consist of 95 males and 74 females. The distribution of the participants per course enrolled is as follows: 10 students in CMSC 100 (Web Programming\footnote{http://www.ics.uplb.edu.ph/courses/ugrad/cmsc/100}), 13 students were enrolled in CMSC 127 (File Processing and Database Systems\footnote{http://www.ics.uplb.edu.ph/courses/ugrad/cmsc/127}), 137 students enrolled in CMSC 128 (Introduction to Software Engineering\footnote{http://www.ics.uplb.edu.ph/courses/ugrad/cmsc/128}), and nine students were enrolled in other various courses whose respective instructors required them to form a group whose collective output becomes their individual outputs. These courses required students to form a group and solve several computer programming tasks throughout the semester. CMSC 100 required each group to design and implement a responsive and immersive web application, usually using the various recent web technologies and tools such as PHP, HTML 5.0, Ajax, and CSS. CMSC 127 required each group to design and implement a real-world database system, often with LGU or private entities as clients and students are free to use their preferred enterprise-like servers such as MariaDB, MySQL, PostgreSQL, MS-SQL Server, and Oracle. CMSC 128  required each group to design, implement and evaluate a computerized solution to a manual processing system (e.g., inventory system, accounting system, etc.). Other courses such as CMSC 198 required the students to work in group in a host company and implement algorithmic solutions to real-world problems. 

\subsection{Survey Questions}

The participants were asked to voluntarily complete a survey form to report their perceptions of the following: 
\begin{enumerate}
\item Degree to which their fellow group members engaged in social loafing; 
\item Personal degree of social loafing; 
\item Individual task visibility; 
\item Individual contribution; 
\item Distributive justice; 
\item Sucker effect; and 
\item Group member dominance. 
\end{enumerate}

All survey questions were adapted from~\citet{George92}, \citet{Piezon08}, and~\citet{Welbourne95} and were the same questions asked in the work of~\citet{Pabico08}. Compared to the 2008 work which was done using the pen-and-paper manner, the responses of the respondents were collected using Google Forms\footnote{https://www.google.com/forms}. With the form, participants were forced to completely reply to the statements presented unlike in the previous where some were not replied to by the respondents.

\section{Results and Discussion}\label{sec:result}

\subsection{Demography of Participants}

\begin{figure*}[hbt]
\centering\epsfig{file=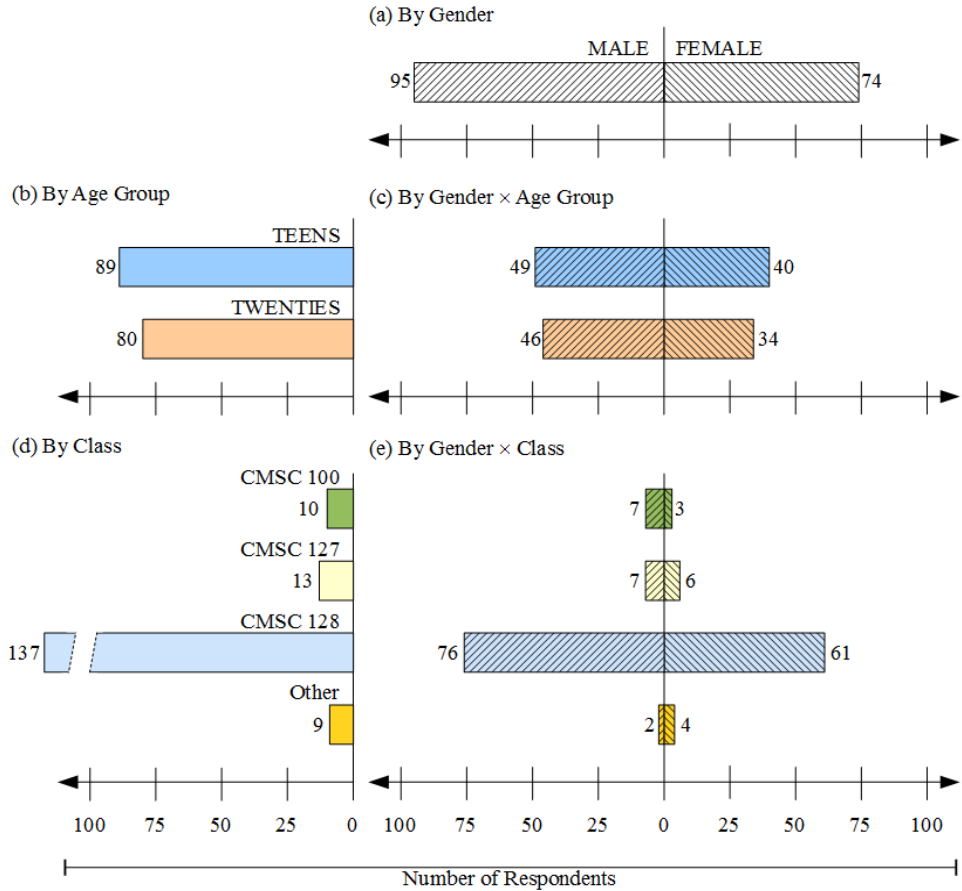, width=5in}
\caption{The descriptive demographic distribution of the respondents according to personal data: (a)~By gender, (b)~by age group, (c)~by gender $\times$ age group, (d)~by class, and (e)~by gender $\times$ class.}\label{fig:1}
\end{figure*}

Figure~\ref{fig:1} shows the descriptive personal demography of the respondents. Male respondents consist 56\% of the respondents with 95 while female consist 44\% with 74 participants (Figure~\ref{fig:1}a). Teens account for 53\% of the respondents with 89 while those who are in twenties consist 47\% with 80 participants (Figure~\ref{fig:1}b). The gender by age group distributions (Figure~\ref{fig:1}c) are as follows: Male teens consist 29\% with 49 respondents, male in twenties are not too far away at 24\% with 40 participants; There were 46 female teens (27\%) and 34 female in their twenties (20\%). Among the classes (Figure~\ref{fig:1}d), CMSC~128 had the most participants at 81\% (137 respondents) while the rest are 10\%, 13\%, and 9\% respectively for CMSC~100 (10 respondents), CMSC~127 (13 respondents), and Other classes (9 respondents). Figure~\ref{fig:1}e show the gender by class distribution, with the males and females in the CMSC~128 class respectively consist of 45\% (76) and 36\% (61) of the total respondents.

\begin{figure*}[hbt]
\centering\epsfig{file=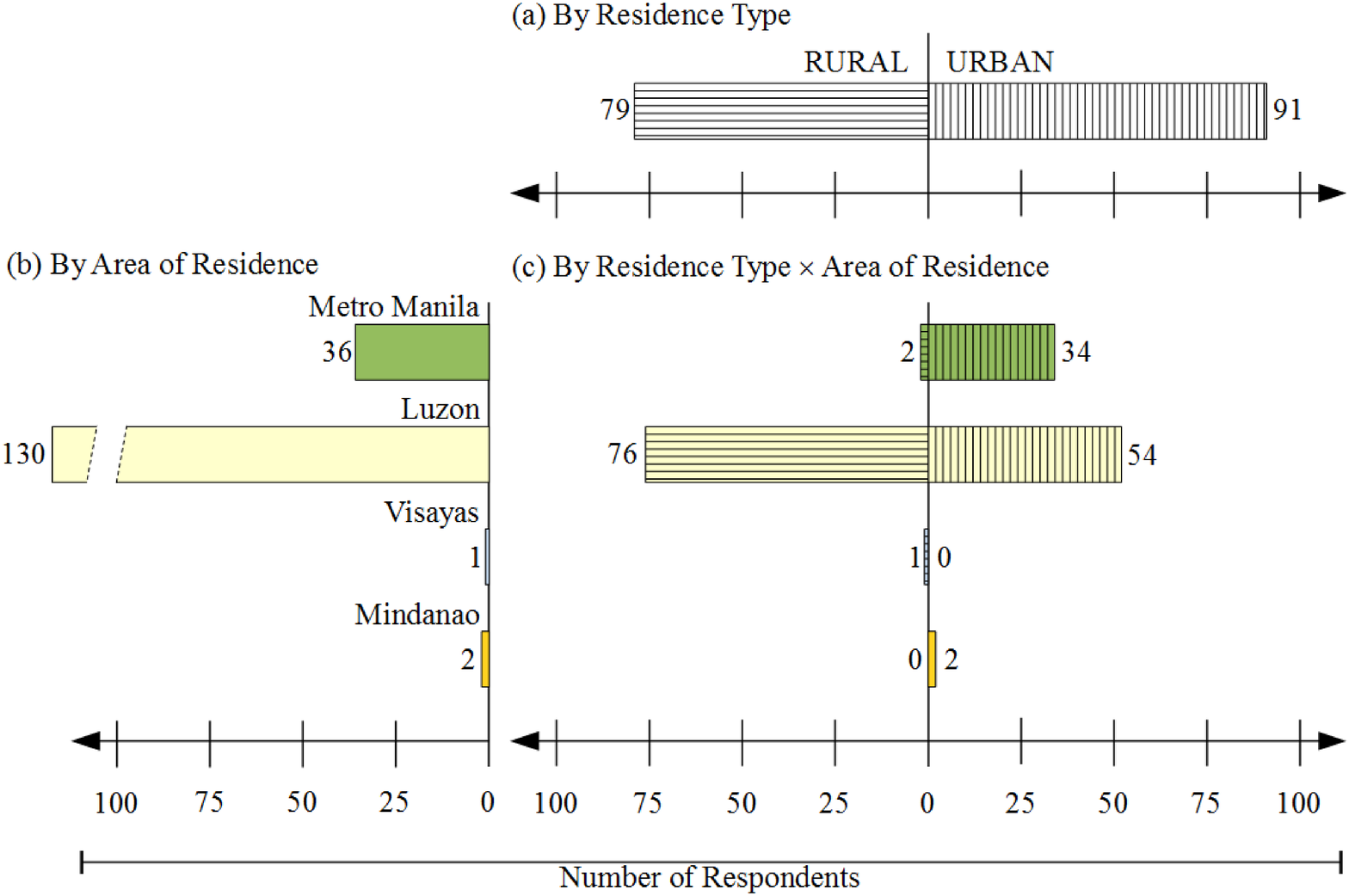, width=5in}
\caption{The descriptive demographic distribution of the respondents according to place of residence data: (a)~By residence type, (b)~by area of origin, and (c)~by residence type $\times$ area of origin.}\label{fig:2}
\end{figure*}

Figure~\ref{fig:2} shows the descriptive demography of the respondents according to their residencial origin. In terms of residence type (Figure~\ref{fig:2}a), 47\% came from the rural areas (79 respondents) while 54\% came from the urban areas (91 participants). We used the four general main islands of Luzon, the Visayas, and Mindanao, and Metro Manila to classify the area of residence of the respondents (Figure~\ref{fig:2}b). Since UPLB serves mainly the Luzon students, it is not surprising the a huge percentage of the respondents from Luzon is recorded at 77\% with 130 students. Coming in second are students from Metro Manila with 21\% (36 respondents). Participants from the Visayas and Mindanao consist of 1\% each at~1 and~2, respectively. The distribution by residence type and area of residence are as follows: 45\% came from the rural Luzon areas (76 participants) while 32\% from the urban Luzon areas (54 participants), excluding Metro Manila. Those coming from the urban Metro Manila consist 20\% of the respondents with 34. It is interesting to note that 1\% of the respondents (2) came from rural Metro Manila, the same percentage as that from rural Visayas and urban Mindanao with 1~and 2~respondents, respectively.

\subsection{Existence of Social Loafing}

\begin{figure*}[hbt]
\centering\epsfig{file=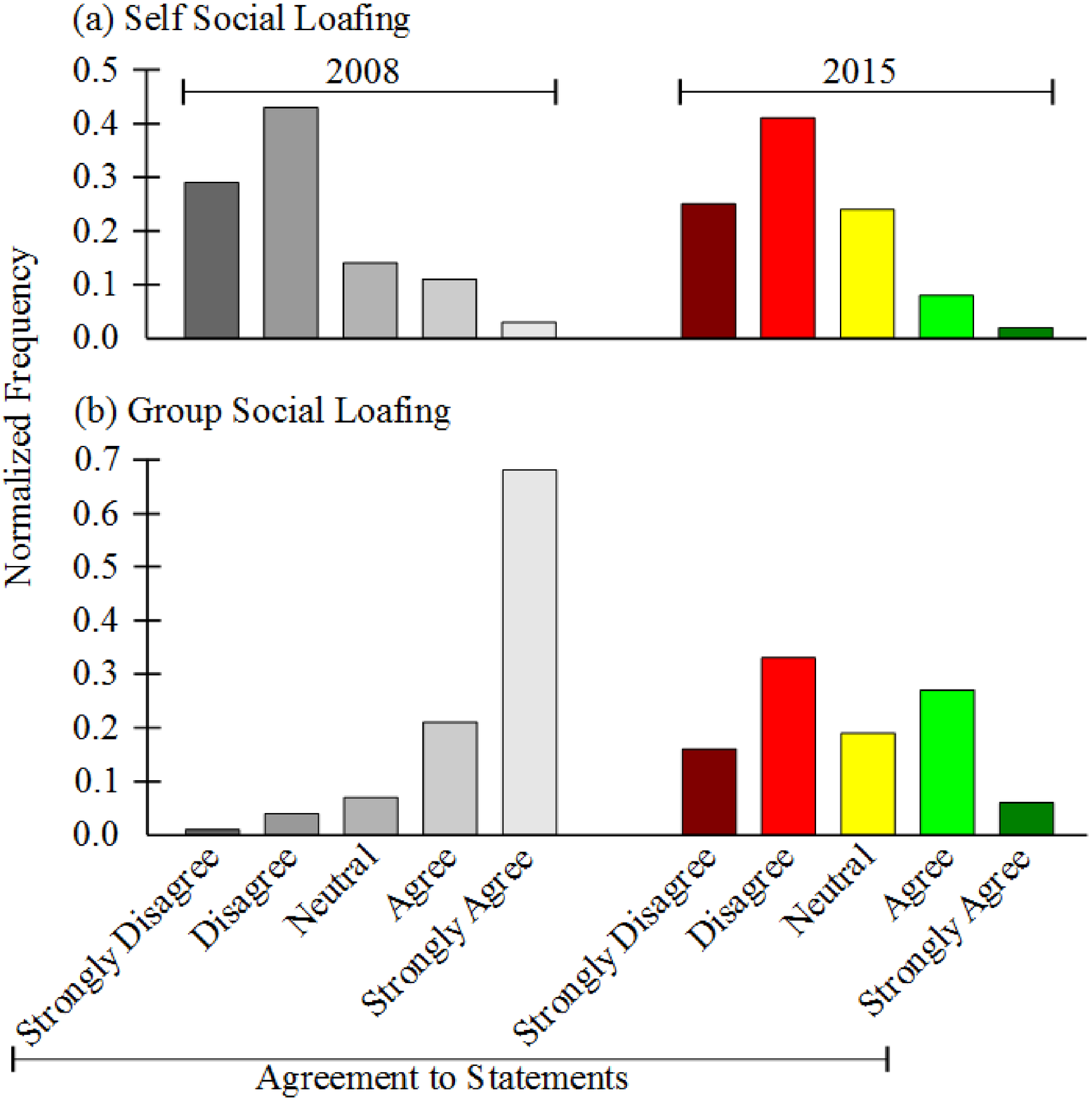, width=4in}
\caption{The respondents agreement that (a)~they were engaged in social loafing and (b)~their perception that their groupmates were engaged in social loafing. The data from seven years ago came from~\citet{Pabico08}.}\label{fig:3}
\end{figure*}

Among the 169 students, about 10\% answered positively relating to their personal engagement in social loafing, while about 66\% answered negatively. The rest (24\%) were not sure whether they engaged in social loafing or not (Figure~\ref{fig:3}a). Compare this result to the work of~\citet{Pabico08} where 14\% of the 239 respondents reported they were engaged in social loafing, while a huge 72\% said that they were not engaged. Those who were not sure consist only of the same number as those who said they were social loafing at 14\%. 

A small portion of the respondents in this study (33\%) reported that they believe their group members were engaged in social loafing, while 49\% said they did not believe so (Figure~\ref{fig:3}b). Those who were not sure consist of 19\% of the respondents. Contrast this result to the work of~\citet{Pabico08} where 89\% reported that they believe their groupmates were social loafing, while only 4\% believed that their groupmates were not. Only 7\% reported neutrality to this issue.

The main goal of our study is to determine whether the perception of social loafing exists in undergraduate computer programming groups. Our results already suggest that social loafing exists (10\% reported they were social loafing as shown in Figure~\ref{fig:3}a and 33\% they perceived their groupmates to be social loafing as shown in Figure~\ref{fig:3}b). Thus, we accept the truthfulness of Hypothesis~\ref{hyp:socloaf}. Although there is a low percentage of self-reported social loafing, it is consistent with the research results of others~\citep{Karau93}. Other studies explained that the students may be unaware that they were social loafing or were just reluctant to admit that they themselves were engaging in social loafing. This observation is specially true to students from a university of known academic activism, such as UPLB, where students pride themselves of individual accomplishments. It is interesting to note that today's students have a relatively higher regards to their groupmates than those who were pooled seven years ago.

\subsection{Perceptions of Precursors}

\begin{figure*}[hbt]
\centering\epsfig{file=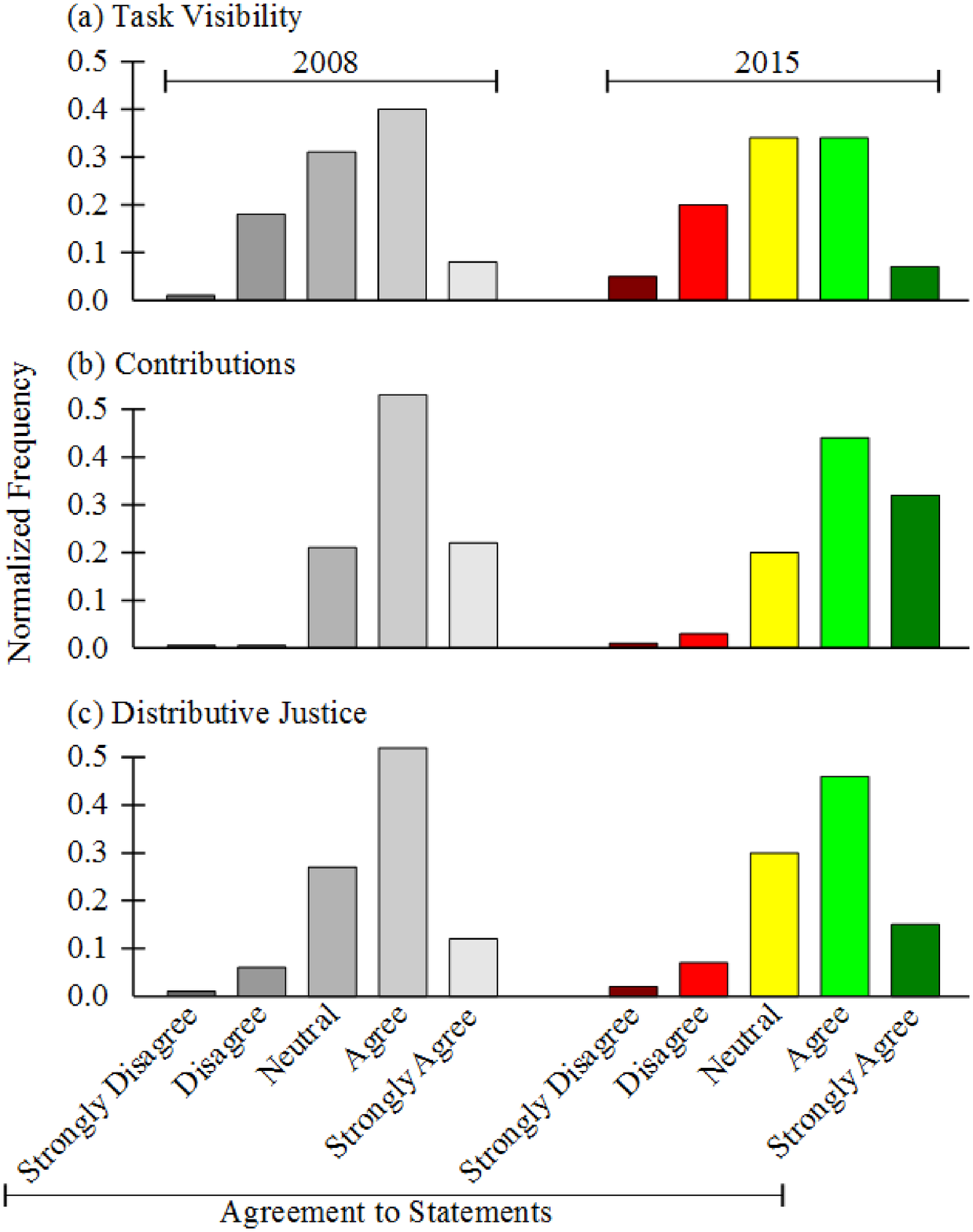, width=4in}
\caption{The respondents agreement to negative precursors to social loafing: (a)~Task visibility, (b)~contributions, and (c)~distributive justice. The data from seven years ago came from~\citet{Pabico08}.}\label{fig:4}
\end{figure*}

Figure~\ref{fig:4} shows the students' response to negative precursors to social loafing: Task visibility, contributions, and distributive justice. The general response of the participants were that of neutrality to that of agreement on the perception that the instructors are tracking their progress to the group's effort, which is generally similar to those who were pooled seven years ago. They were also in agreement to the statements that their contributions are important to the success of the project and that their instructors will reward them with benefits justly.

\begin{figure*}[hbt]
\centering\epsfig{file=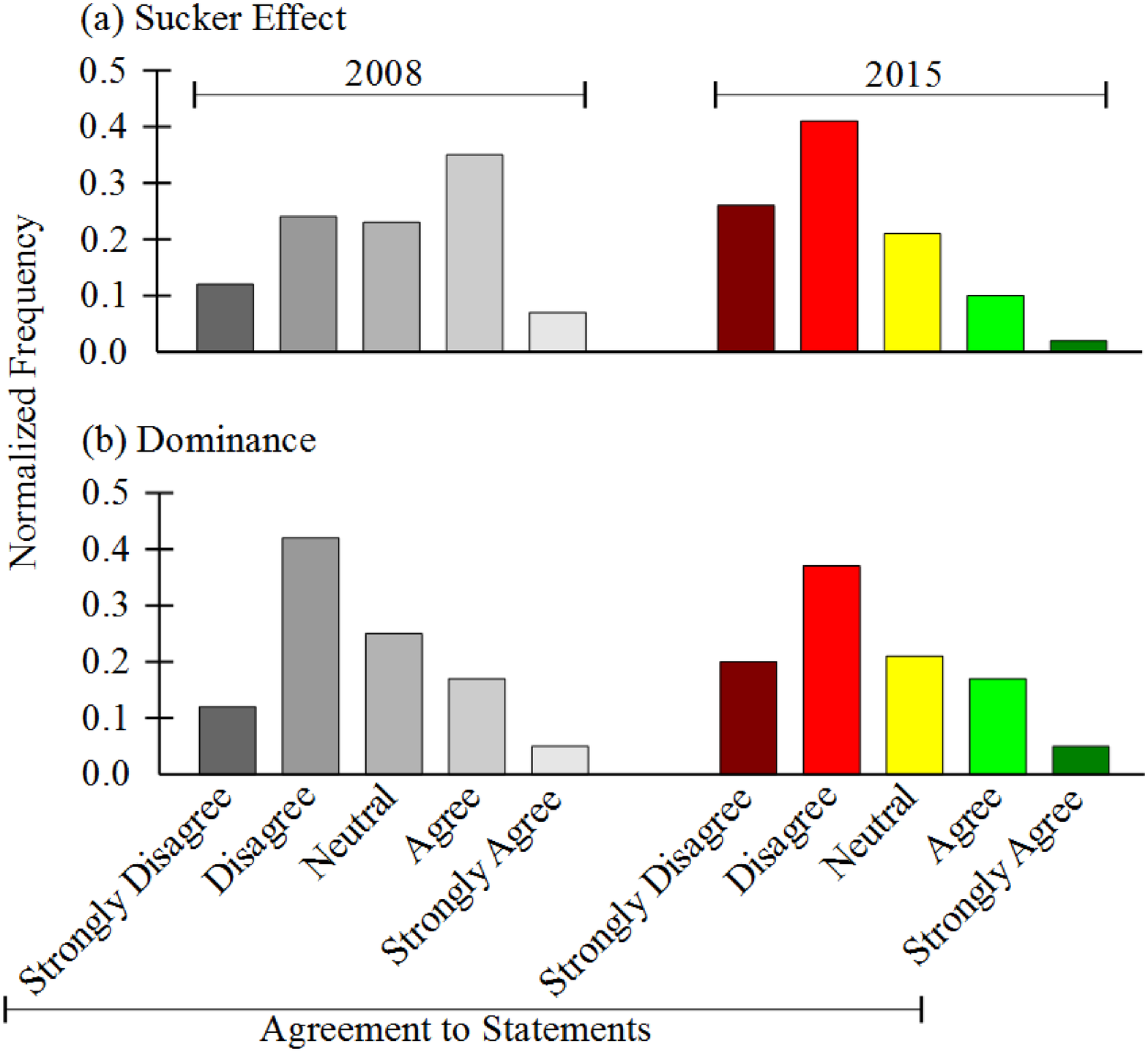, width=4in}
\caption{The respondents agreement to positive precursors to social loafing: (a)~Sucker effect, and (b)~dominance and agression. The data from seven years ago came from~\citet{Pabico08}.}\label{fig:5}
\end{figure*}

Figure~\ref{fig:5} shows the students' response to positive precursors to social loafing: Sucker effect and dominance. Today's students generally disagreed with their perception that both sucker effect and dominance exist in their groups. However, students who were asked to their respective agreements to the same statements seven years ago generally agreed to the perception of sucker effect but disagreed to the perception of dominance in their groups.

\subsection{Social Loafing}

Tables~\ref{tab:cor-all} shows the result of the analysis with respect to the responses of all participants: The correlations of the precursors to the respondents' perception on the social loafing of their groupmates and their own acknowledgment to social loaf. The precursors Contributions is negatively correlated to self social loafing while Sucker Effect and Dominance are positive correlated to self social loafing with respective significant Pearson correlations of -0.49, 0.63, and 0.37. Task Visibility and Distributive Justice each had negative coefficients but were found to be not significantly different from zero at $\alpha=0.05$.

\begin{table}[hbt]
\caption{Correlation analysis of the response of all participants, where at each coefficient the superscript {\em ns} means not significant, * means significant at $\alpha=0.05$, and ** means significant at $\alpha=0.01$.}\label{tab:cor-all}
\centering\begin{tabular}{lcc}
\hline\hline
\multirow{2}{*}{Precursor} & \multicolumn{2}{c}{Social Loafing}\\
\cline{2-3}
                           & Group & Self \\
\hline
Task Visibility & -0.20\textsuperscript{*}  & -0.07\textsuperscript{ns} \\
Contributions   & -0.10\textsuperscript{ns} & -0.49\textsuperscript{**} \\
Sucker Effect   &  0.15\textsuperscript{ns} &  0.63\textsuperscript{**} \\
Distributive Justice & -0.09\textsuperscript{ns} & -0.11\textsuperscript{ns}\\
Dominance       &  0.20\textsuperscript{**} &  0.37\textsuperscript{**}\\
\hline\hline
\end{tabular}
\end{table}

The precursor Task Visibility was found to be negatively correlated to Group Social Loafing, while Dominance was positively correlated with respective significant coefficients of -0.20 and 0.20 at $\alpha=0.05$. These results can be abstracted as follows:
\begin{eqnarray}
 \PERCEPT_\TASKVIS & \propto & -\PERCEPT_\GROUPSL\\\label{eqn:taskvis}
 \PERCEPT_\CONTRIB & \propto & -\PERCEPT_\SELFSL\\\label{eqn:contrib}
 \PERCEPT_\SUCKER  & \propto & \PERCEPT_\SELFSL\\\label{eqn:sucker}
 \PERCEPT_\DOMINANCE & \propto & \PERCEPT_\SELFSL + \PERCEPT_\GROUPSL\label{eqn:dominance}
\end{eqnarray}

Equations~\ref{eqn:taskvis}, ~\ref{eqn:contrib}, ~\ref{eqn:sucker}, and~\ref{eqn:dominance} respectively provide support to Hypotheses~\ref{hyp:taskvis}, \ref{hyp:contrib}, \ref{hyp:sucker}, and~\ref{hyp:dominance}.

Equation~\ref{eqn:taskvis} suggests that the perception of increased individual task visibility decreases the perceived social loafing by one's groupmate. Equation~\ref{eqn:contrib} says that the perception of increased contribution to group's effort decreases the social loafing by one's self. Equation~\ref{eqn:sucker} means that an increase in the perception of sucker effect increases the social loafing of one's self, while Equation~\ref{eqn:dominance} means an increase in the perception of dominance and aggression in the group increases both the perception of social loafing by a groupmate and by one's self.

Table~\ref{tab:cor-gender} shows the correlations by gender of the precursors to the respondents' perception on the social loafing of their groupmates and their own acknowledgment to social loaf. Among female respondents, Contributions is negatively correlated to self social-loafing while Sucker Effect and Dominance are positively correlated with significant coefficients of -0.49, 0.12, and 0.54, respectively, at $\alpha=0.05$. Task Visibility is negatively correlated to Group Social Loafing with coefficient of -0.25. As in Self Social Loafing, Sucker Effect and Dominance are both positively correlated with respective coefficients 0.26 and 0.35 ($\alpha=0.05$) with Group Social Loafing.

\begin{table}[hbt]
\caption{Correlation analysis of the response of participants by gender, where at each coefficient the superscript {\em ns} means not significant, * means significant at $\alpha=0.05$, and ** means significant at $\alpha=0.01$.}\label{tab:cor-gender}
\centering\begin{tabular}{lcc}
\hline\hline
\multirow{2}{*}{Precursor} & \multicolumn{2}{c}{Social Loafing}\\
\cline{2-3}
                           & Group & Self\\
\hline
                & \multicolumn{2}{c}{Female}\\
\cline{2-3}
Task Visibility & -0.25\textsuperscript{*}  & -0.19\textsuperscript{ns} \\
Contributions   & -0.16\textsuperscript{ns} & -0.49\textsuperscript{**} \\
Sucker Effect   &  0.26\textsuperscript{*}  &  0.74\textsuperscript{**} \\
Distributive Justice & -0.22\textsuperscript{ns} & -0.12\textsuperscript{ns}\\
Dominance       &  0.35\textsuperscript{**} &  0.55\textsuperscript{**}\\
\hline
                & \multicolumn{2}{c}{Male}\\
\cline{2-3}
Task Visibility & -0.12\textsuperscript{ns} &  0.04\textsuperscript{ns} \\
Contributions   & -0.05\textsuperscript{ns} & -0.49\textsuperscript{**} \\
Sucker Effect   &  0.03\textsuperscript{ns} &  0.56\textsuperscript{**} \\
Distributive Justice & 0.03\textsuperscript{ns} & -0.12\textsuperscript{*}\\
Dominance       &  0.03\textsuperscript{ns} &  0.24\textsuperscript{**}\\
\hline\hline
\end{tabular}
\end{table}

The male participants perceived Contributions (-0.49) and Distributive Justice (-0.12) negatively with Self Social Loafing, while Sucker Effect (0.56) and Dominance (0.24) are positively perceived with Self Social Loafing. All correlations are significantly different from zero at $\alpha=0.05$. Male respondents do not perceive social loafing among their groupmates.

The abstraction of the above results by gender are as follows:

\begin{eqnarray}
 \PERCEPT_\TASKVIS & \propto & -\PERCEPT_\GROUPSL^\FEMALE\\\label{eqn:taskvis-gender}
 \PERCEPT_\CONTRIB & \propto & -\PERCEPT_\SELFSL^{\FEMALE+\MALE}\\\label{eqn:contrib-gender}
 \PERCEPT_\SUCKER  & \propto & \PERCEPT_\SELFSL^{\FEMALE+\MALE} + \PERCEPT_\GROUPSL^\FEMALE\\\label{eqn:sucker-gender}
 \PERCEPT_\JUSTICE & \propto & -\PERCEPT_\SELFSL^\MALE\\\label{eqn:justice}
 \PERCEPT_\DOMINANCE & \propto & \PERCEPT_\SELFSL^{\FEMALE+\MALE} + \PERCEPT_\GROUPSL^\FEMALE\label{eqn:dominance-gender}
\end{eqnarray}

Equations~\ref{eqn:taskvis-gender}, \ref{eqn:contrib-gender}, \ref{eqn:sucker-gender}, and~\ref{eqn:dominance-gender} already confim the formulations in Equations~\ref{eqn:taskvis}, \ref{eqn:contrib}, \ref{eqn:sucker}, and~\ref{eqn:dominance}, respectively. The new result here is that of Equation~\ref{eqn:justice} which confirms Hypothesis~\ref{hyp:justice}.

All the results that we have so far presented and discussed already confirm all our Hypotheses. Thus, for reasons of consciseness, we do not need to present all other remaining detailed correlation analyses by age, class, type of residence and area of residence. We will, however, present all these in the future. Note here that all our results agree with the findings of others. In particular:
\begin{enumerate}
\item With~\citet{Liden04} who suggested that non-recognition of an individual's input often leads to social loafing. On the other hand, a positive perception of recognition of one's contribution decreases the occurrence of social loafing. 
\item With~\citet{Liden04}, whose work suggests that the positive perception of the distribution of grades among members will decrease the occurrence of social loafing. This means that ensuring that the group members understand the procedures behind the grade distribution can have a positive influence on their behavior in the group. If a group member either misunderstand or perceive inequitable grade distribution, she may engage in social loafing in order to balance the perceived reward-per-effort ratio. 
\end{enumerate}

\section{Conclusion}\label{sec:conclude}

We studied the existence of social loafing among members of undergraduate computer programming groups who were enrolled in various separate classses in two semesters at UPLB. We asked the students to voluntarily answer an online survey question that will determine the occurrence of several precursor behaviors to social loafing. We found out that social loafing exists among the group members. Based on the correlation analysis that~\citet{Pabico08} conducted seven years ago, the following relationships were established: 
\begin{enumerate}
\item There is a negative correlation between the perceived task visibility and the perceived social loafing. 
\item There is a negative correlation between the perceived individual contribution and the perceived social loafing. 
\item There is a negative correlation between the perceived distributive justice and the perceived social loafing. 
\item There is a positive correlation between the perceived dominance and the perceived social loafing. There is also a positive correlation between the perceived dominance and sucker effect. 
\end{enumerate}

These same results are re-established in this research effort. The results of our study provide evidence that the precursor behaviors to social loafing exist and  still persist in undergraduate computer programming groups at UPLB. This implies that instructors and course designers, even though already considered the recommendations offered by~\citet{Pabico08} to consider social loafing when designing groups, must continue to innovate to adapt to the varying needs of today's students. Aside from adaptive mechanisms by students to innovations introduced by instructors and course designers, social loafing may be a needed behavior to continue to survive the so called {\em academic jungle}. We believe, as what~\citet{Pabico08} also believed, that social loafing may not be at all bad under some circumstances~\citep{Jackson85}. Because of reduced effort, social loafing may result in reduced stress for the student, and thereby improved her performance later. We will continue to echo the use of currently available technologies to improve the tracking of a student's inputs. Aside from the use of the Concurrent Version System to track changes to computer programming codes, we suggest the use of the cloud-based GitHub which not only offers desktop-based applications but smart-device-based ones, too.

\section*{Acknowledgements}

This research effort is funded by the Institute of Computer Science Core Fund for AY 2014-2015 and AY 2015-2016. We thank the 169 students who unselfishly donated their time in honestly responding to the questions asked in the online questionaire. These students embody the true {\em iskolar ng bayan para sa bayan}.

\bibliography{paper1}
\bibliographystyle{plainnat}
\end{document}